# Microservices, Continuous Architecture, and Technical Debt Interest: An Empirical Study

Valentina Lenarduzzi, Davide Taibi
Tampere University of Technology
Tampere, Finland
{valentina.lenarduzzi; davide.taibi}@tut.fi

Continuous Architecture (CA) is an approach that supports companies in decreasing the time between deliveries. Erder and Pureur define CA as an approach based on a toolbox that helps companies make the right architectural decisions [1]. The CA principle recommends delaying design decisions until they are absolutely necessary [1]. The developed system should be architected to enable changes, leveraging "The Power of Small". Moreover, the systems should be architected with a special focus on the build, test, and deploy phases. Finally, the CA principle also suggests following Conway's law [2], modeling the organization of the development teams after the design of the system they are working on.

Migration to microservices is one of the most common situations when companies adopt continuous architecting processes [4]. Companies commonly adopt an initial migration strategy to extract some components from the monolithic system as microservices, making use of simplified microservices patterns [5][4]. As an example, companies commonly directly connect the microservices to the legacy monolithic system and do not adopt any message bus at the beginning. When the system starts to grow in complexity, they usually start re-architecting their systems, considering different architectural patterns. Some companies introduce API gateway patterns to simplify the management and load balancing of the different services, while others also consider a lightweight message bus [4][5][6]. All these architectural changes require deep refactoring of the system, thereby increasing the risk of new faults being introduced.

However, software development is commonly driven by the generation of new features and developers often postpone some refactoring activities or adopt temporary solutions in order to deliver new features on time. Each postponed activity can be considered as part of the technical debt (TD) [8]. Moreover, the postponed activities and the temporary solution are commonly developed within a very short period of time, which dramatically increases the risk of faults [3] as well as the interest on the TD.

In this paper, we report the preliminary results of work in progress, where we monitored the TD of an SME (SMEs = small and medium enterprises) that adopted a CA approach while migrating a legacy monolithic system to an ecosystem of microservices. To the best of our knowledge, no studies exist on the impact of postponed activities on the TD, especially in the context of CA and microservices. This work will help companies understand how TD grows and changes over time while at the same time opening up new avenues for future research on the analysis of TD interest in continuous architecting processes.

## I. THE CASE STUDY

In this preliminary case study, we aimed to compare the technical debt (TD) and its trend in a project that migrated to microservices. According to our expectations, we formulated two research questions (**RQs**):

**RQ1**: Is the technical debt of a monolithic system higher than the technical debt of the same system developed with a microservices architectural style?

**RQ2**: Is the technical debt of a monolithic system growing with the same trend as a microservices-based system?

In order to answer our research questions, we monitored the development of a document management system developed by a local SME. The system is being developed in Java, is deployed on the Microsoft Azure Cloud, and is delivered to the customers as a web application, together with a desktop application to support document uploading and synchronization. The monolithic system was composed of 280K lines of code and had been developed for more than twelve years. The company decided to migrate to microservices in order to make maintenance easier by separating each business process better and to reduce the need for synchronization between the two development teams. Two teams, one composed of four developers and another one composed of five developers, started the migration in March 2017. In January 2018, the teams extracted four business processes into four independent microservices. Two of the six microservices were deeply re-architected after the first release: In one service, the MySQL database was replaced by a MongoDB; the other service was re-written in Python, so as to benefit from a set of machine learning libraries not available in Java.

### A. Study Preparation, Execution, and Analysis

We first accessed the SCM (Gitlab) and the issue tracker (Jira) adopted by the company. We analyzed the TD of each commit over the two years prior to the migration and during the migration. The TD was analyzed by means of SonarQube[9], the most commonly adopted static code analysis tool, by applying the standard quality profile. The preliminary analysis was carried out by comparing the trend of the TD series.

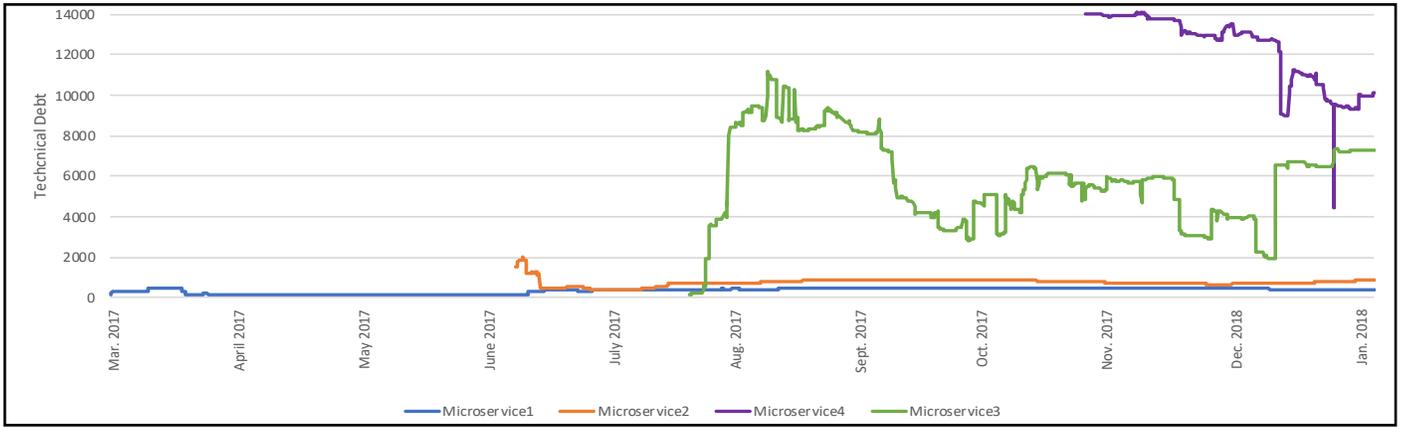

Figure 1: Technical Debt Evolution of the Four Microservices

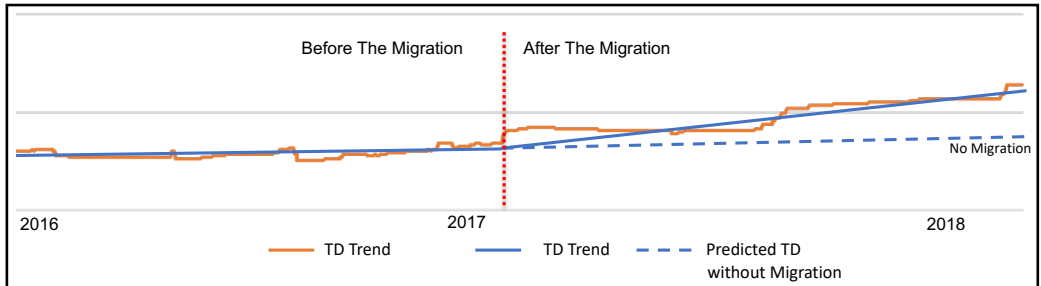

Figure 2: Overall Technical Debt Evolution of the whole System

*B. Preliminary Results*

The company started to develop microservices gradually. They began with the development of a single service, extracting a business process they considered to have low risk.

**The technical debt of a monolithic system before migration is always lower than the technical debt of the system after migration**. Technical debt tends to be more stable and to increase slower in each microservice compared to the whole system's technical debt but the total TD (sum of the TD of the monolithic system and all the microservices) grows faster compared to the TD growth before the migration. However, TD in each microservice tends to decrease after a relatively short time. Since the vast majority of business processes are still in the monolithic system, we expect the technical debt of the whole system to be lower compared with the technical debt of the monolithic system after the migration of all the processes.

*C. Discussion and Threats to Validity*

The results of this study show that technical debt tends to grow slower in microservices compared to monolithic systems. The findings of this work are based on only one preliminary case study carried out with a single company. Moreover, we collected the technical debt using the model provided by SonarQube. Therefore, different tools and different approaches might provide different results.

## II. NEXT STEPS

We are currently working on the definition of the interest on the technical debt while developing microservices. Every time a technical issue is postponed, developers estimate the time they would need to develop it. In this case, we are considering all possible issues, including both SonarQube issues, architectural decisions, and any other possible postponed activity (e.g., adoption of another database or use of a different architectural pattern). When the issue is being developed, we are tracking the actual time taken by the developer to implement it, so as to understand how long the postponed activity took, compared to the time it would have taken had it not been postponed.

Our next goal is to understand how long different activities could be postponed before the benefit of postponing an activity is canceled out by the increased effort needed to refactor it. As an example, if an activity has an interest equal to zero (i.e., if the development/refactoring effort does not increase if postponed), it can be postponed until it is needed, whereas if an activity has a monthly interest of 10% (i.e., 10% extra interest per month), it should be refactored as soon as possible before getting too expensive. The interest will be identified both with interviews to developers and with mining software repositories methods, following the approach adopted in [7].

## III. CONCLUSION

Continuous architecting recommends postponing decisions until they are absolutely necessary. This approach is especially applied in the context of migration of microservices. In this work, we conducted a preliminary case study to understand the trend of technical debt monitoring of an SME that migrated its monolithic system to microservices. The result is that **the total** amount of **technical debt grows much faster in a microservices-based system**, probably due to the large number of postponed refactoring activities. The migration to microservice initially increase the amount of technical debt and the more the activities are delayed, the more interest will be accrued.